\documentclass[11pt]{article}
\usepackage{amssymb,amsmath}
\usepackage{graphicx}
\usepackage{tabbert} 

\newcommand{\hu}{\hat{u}}
\newcommand{\hT}{\hat{T}}
\newcommand{\hS}{\hat{S}}

\allowdisplaybreaks[2]

\title{An alternative to the Allen-Cahn phase field model for
  interfaces in solids -- numerical efficiency}

\author{
Hans-Dieter Alber\addtocounter{footnote}{1}\footnote{alber@mathematik.tu-darmstadt.de}
\\
{\small Fachbereich Mathematik, Technische Universit\"at Darmstadt} \\ 
{\small Schlossgartenstr. 7, 64289 Darmstadt, Germany}
}

\date{}
\begin{document}
\maketitle

\begin{abstract}
  The derivation of the Allen-Cahn and Cahn-Hilliard equations is
  based on the Clausius-Duhem inequality. This is not a derivation in
  the strict sense of the word, since other phase field equations can
  be fomulated satisfying this inequality. Motivated by the form of
  sharp interface problems, we formulate such an alternative equation
  and compare the properties of the models for the evolution of phase
  interfaces in solids, which consist of the elasticity equations and
  the Allen-Cahn equation or the alternative equation. We find that
  numerical simulations of phase interfaces with small interface
  energy based on the alternative model are more effective then
  simulations based on the Allen-Cahn model.
\end{abstract}

\begin{center}
Dedicated to the memory of Krzysztof Wilma{\'n}ski
\end{center}

\section{Introduction}\label{S1}

The phase field approach is used to model the evolution of phase
interfaces in many different materials and accordingly the resulting
models differ widely. However, in spite of all the differences the
evolution equations for the order parameter $S$ in the models is
almost always formulated by the standard approach to set the time
derivative of the order parameter equal to a suitable function of the
functional derivative of the Ginzburg-Landau free energy with respect
to $S$, which leads to an Allen-Cahn type equation, or equal to the
divergence of a suitable function of the gradient of this functional
derivative, which leads to a Cahn-Hilliard type equation. Often this
function is chosen to be linear. For a thorough discussion of this
approach to formulate material models with the Allen-Cahn und
Cahn-Hilliard equation we refer to \cite{BS1998}.

The leading idea behind these approaches to formulate the evolution
equation is that in both cases for the resulting model the
Clausius-Duhem inequality is guaranteed to hold. Yet, there are other
possibilities to choose the evolution equation such that this
inequality holds. Therefore the question arises whether the standard
approach is always the best or whether there are situations where
other choices of the evolution equation for the order parameter lead
to better results. 

Of course, this question can only be discussed at a concrete example
of an alternative phase field equation in a concrete mathematical
material model. We consider here the prototypic model for the
evolution of phase interfaces in solids, neglecting temperature
effects, which consists of the elasticity equations coupled to an
evolution equation for $S$. For this evolution equation one usually
inserts the Allen-Cahn equation. We formulate here an alternative
phase field equation and compare the mathematical properties of the
two different models, which are obtained when we use the Allen-Cahn
equation or the alternative equation as the evolution equation.  Our
main result is that simulations of phase interfaces in solids, which
have small or vanishing interface energy density, are numerically more
effective when the alternative equation is used instead of the
Allen-Cahn equation.

We stress that the alternative phase field equation can replace the
Allen-Cahn equation in other models. The properties of the resulting
models have as yet to be investigated.  

This paper is based on our investigations of phase field equations in
the articles \cite{Al00} -- \cite{AC2015}. It summarizes in particular
the results obtained in \cite{JElast2013} and \cite{AC2015}, but adds
also some new considerations.

\section{ The Clausius-Duhem inequality and the Allen-Cahn
  equation}\label{S2}

To formulate the alternative phase field equation, we must know the
form of the Ginzburg-Landau free energy, which appears in the
Clausius-Duhem inequality. Therefore we first introduce the physical
situation and the elasticity equations, from which the form of the
Ginzburg-Landau free energy results. 

Let $\Om \subseteq \R^3$ be an open bounded set, which represents a
solid body. We assume that the atoms of the material can be arranged
in crystal lattices of two different types. The crystal type present
at a material point $x \in \Om$ at time $t$ is indicated by the order
parameter. The value $S(t,x) = 0$ means that type one is present,
$S(t,x) = 1$ indicates that type two is present.  The sets of points
\[
\gm(t) = \{ x \in \Om \mid S(t,x) = 0 \}, \qquad \gm'(t) = \{x \in \Om
\mid S(t,x) = 1 \}, 
\]
where crystal type one or crystal type two is present, respectively,
are called phase~$1$ or phase~$2$ of the material at time $t$,
respectively. Let $u(t,x) \in \R^3$ denote the displacement of the
material point $x$ at time $t$ and let 
\[
\ve(\na_x u) = \frac12 \big( \na_x u + (\na_x u)^T \big) \in \ES^3
\]  
be the linear strain tensor, where $\ES^3$ denotes the set of
symmetric $3\ti3$--matrices. We assume that only small displacements
occur and we consider a quasistatic model. This means that for every
given time $t$ the displacement $x \mapsto u(t,x)$ and the Cauchy
stress tensor $x \mapsto T(t,x) \in \ES^3$ must solve the boundary
value problem of linear elasticity posed in the domain $\Om$, which is
given by 
\begin{eqnarray}
-\div_x\, T &=& {\sf b},  \label{E2.1}
\\
T &=& D\big(\ve(\na_x u ) - \ov{\ve}S \big),\label{E2.2}
\\
u(t,x) &=& {\sf U}(t,x), \qquad x \in \pa \Om, \label{E2.3}
\end{eqnarray}
where $\ov{\ve} \in \ES^3$ is the given transformation strain, where
$D:\ES \ra \ES$ is the elasticity tensor, a linear, symmetric,
positive definite mapping, and where ${\sf b}(t,x), {\sf U}(t,x) \in
\R^3$ denote the given volume force and boundary displacement. By
\eq{2.2}, the material is stree free in phase one if $\ve(\na_x u)$ is
equal to zero, and in phase two if $\ve(\na_x u)$ is equal to the
transformation strain $\ov{\ve}$.

To close the system of model equations, we need an evolution equation
for $S$. To formulate it, note that according to \eq{2.2}, the stored
elastic energy is 
\begin{equation}\label{E2.4}
{\sf W}\big(\ve(\na_x u),S\big) = \frac12 \Big(D \big( \ve(\na_x u) -
  \ov{\ve} S \big) \Big) : \big( \ve(\na_x u) - \ov{\ve}S \big),
\end{equation}
which leads to the Ginzburg-Landau free energy  
\begin{equation}\label{E2.5}
\psi^*\big(\ve(\na_x u),S,\na_x S\big) =  {\sf W}\big(\ve(\na_x
u),S\big) + \hat{\psi}(S) + \frac12 |\na_x S|^2.  
\end{equation}
where $\hat{\psi}:\R \ra \R$ is a double well potential satisfying 
\begin{equation}\label{E2.5a}
\hat{\psi}(0) = \hat{\psi}(1) = 0, \qquad \hat{\psi}(r) > 0, \mbox{
  for }  0 < r < 1. 
\end{equation}
The second law of thermodynamics requires that there is a flow of the
free energy $q \big( u, u_t, \ve(\na_x u), S, S_t, \na_x S \big)$,
such that the Clausius-Duhem inequality
\begin{equation}\label{E2.6}
\frac{\pa}{\pa t} \psi^* + \div_x q \leq {\sf b}\cdot u_t
\end{equation}
holds for all solutions $(u,T,S)$ of the model equations. We use the
flow 
\begin{equation}\label{E2.7}
q = - T u_t - S_t \na_x S . 
\end{equation}
If we insert \eq{2.5} and \eq{2.7} into \eq{2.6} and note \eq{2.1} and
the equation $\pa_{(\na_x u)}{\sf W} = T$, which follows from
\eq{2.4}, \eq{2.2}, then we obtain by a short computation that
\begin{multline}\label{E2.8}
0 \geq \frac{\pa}{\pa t} \psi^* + \div_x q - {\sf b}\cdot u_t 
  = \pa_{(\na_x u)}{\sf W} : \na_x u_t + \big( \pa_S {\sf W} +
  \hat{\psi}'(S) \big)S_t + \na_x S \cdot \na_x S_t 
\\
- \div_x (T u_t) - \div_x (S_t \na_x S) - {\sf b}\cdot u_t   
  = ( \pa_S {\sf W} + \hat{\psi}'(S) - \Da_x S) S_t\,. 
\end{multline}
The Clausius-Duhem inequality \eq{2.6} is therefore satisfied, if the
evolution equation for $S$ guarantees that the right hand side of
\eq{2.8} is non-positive. The simplest possibility to obtain this is
to set 
\begin{equation}\label{E2.9}
\pa_t S = -f \Big( \pa_S {\sf W}\big(\ve(\na_x u),S\big) +
\hat{\psi}'(S) - \Da_x S \Big),
\end{equation}
with a function $f:\R \ra \R$ satisfying $r \cdot f(r) \geq 0$. If for
$f$ the linear function $f(r) = cr$ is chosen with a positive constant
$c$, then the Allen-Cahn equation results.

\eq{2.1}, \eq{2.2}, \eq{2.9} form a closed system of partial
differential equations. The standard phase field model for the
evolution of phase interfaces consists of this system, combined with
the boundary condition \eq{2.3} and an initial condition for $S$.

\section{Formulation of an alternative phase field equation}

By the inequality \eq{2.8}, the expression 
\begin{equation}\label{E3.1}
{\cal F} = \pa_S {\sf W} + \hat{\psi}'(S) - \Da_x S
\end{equation}
and the time derivative $S_t$ must have opposite signs, which means
that the value of $S_t$ at $(t,x)$ cannot be independent of the value
${\cal F}(t,x)$. Instead, there must be a functional relation between
both values. Of course, this does not mean that $S_t$ must depend on
${\cal F}$ alone as in the ansatz \eq{2.9}, it can depend on
additional variables as well. The question arises, on which other
variables $S_t$ should depend.

To discuss this question we start from the usual physical
interpretation of the observation, that there must be a functional
relation between $S_t$ and ${\cal F}$. The interpretation is that
${\cal F}$ is a configurational force, which drives the time evolution
of the order parameter $S$. This interpretation is used as an
additional justification for the equation \eq{2.9}, which we write in
the short form
\begin{equation}\label{E3.2}
S_t(t,x) = -f\big({\cal F}(t,x)\big).
\end{equation}
What one wants to have is that the variation of the order parameter
$S$ is confined to a narrow diffuse interface, which moves with a
propagation speed, which is a linear or nonlinear function of the
configurational force ${\cal F}$. In fact, standard sharp interface models
contain an equation, which prescribes the propagation speed of the
interface as a function of the configurational driving force. This
equation is called kinetic relation. We extend the meaning of this
notation also to phase field models.

In a standard sharp interface model the kinetic relation can therefore
be explicitly read off from the model equations. It would be of
interest to have a phase field model, where the kinetic relation can
also be read off directly from the form of the model equations. For
the phase field equation \eq{3.2} this is not possible.  Instead, the
kinetic relation is a hidden property of this equation, which must be
determined by a very technical asymptotic analysis of this equation.

Our goal is therefore to formulate a phase field equation, for which
the Clausisus-Duhem inequality \eq{2.6} is satisfied, and which allows
to read off the kinetic relation directly from the form of the
equation. To formulate such an equation, assume that $S$ is an order
parameter, whose transition from $0$ to $1$ defines a diffuse phase
interface moving in time. We say that the speed of the diffuse
interface at $(t,x_0)$ is equal to the normal speed $s(t,x_0)$ of the
level set $\Gm_c(t) = \{ x \in \Om \mid S(t,x) = c \}$, which contains
$x_0$. The normal speed of $\Gm_c(t)$ at $x \in \Gm_c(t)$ can be
defined as follows: If $\tilde{t} \mapsto x(\tilde{t}) \in \R^3$ is a
function defined for all $\tilde{t}$ from a neighborhood of $t$ and if
$x(\tilde{t}) \in \Gm_c(\tilde{t})$ holds for all $\tilde{t}$, then
the normal speed $s(t,x)$ of $\Gm_c(t)$ at $ x = x(t) \in \Gm_c(t)$
is the component of the velocoity $x'(t)$ in the direction of the unit
normal vector $n(t,x)$ to $\Gm_c(t)$ at $x$.  Since $n(t,x) =
\frac{\na_x S(t,x)}{|\na_x S(t,x)|}$, we obtain
\begin{equation}\label{E3.3}
s(t,x(t)) = \frac{{\rm d}x(t)}{{\rm d}t} \cdot \frac{\na_x
  S(t,x(t))}{|\na_x S(t,x(t))|}.
\end{equation}
The function $t \mapsto x(t)$ satisfies $x(t) \in \Gm_c(t)$ if
and only if $t \mapsto S\big(t,x (t)\big)= c$ holds, and this last
equation holds if and only if for a fixed time $t_0$ the function $x(t)$
satisfies the initial value problem  
\begin{multline*}
0= \frac{d}{dt} S\big(t,x(t)\big)=S_t\big(t,x(t)\big) + \frac{dx(t)}{dt}
 \cdot \na_x S\big(t,x(t)\big) 
\\  
= S_t\big(t,x(t)\big) + s\big(t,x(t)\big)|\na_x S\big(t,x(t)\big)|,
\qquad x(t_0) \in \Gm_c(t_0),   
\end{multline*}
with $s$ defined by \eq{3.3}. From this we conclude that if $t_1 <
t_2$ are given times and if $s:[t_1,t_2]\ti \Om \ra \R$ is a given
function, then $S$ satisfies the partial differential 
equation  
\begin{equation}\label{E3.4}
S_t + s |\na_x S| = 0 
\end{equation}
in the domain $[t_1,t_2] \ti \Om$, if and only if every level set
$\Gm_c(t)$ moves with normal speed $s(t,x)$ at $x \in \Gm_c(t)$.

This suggests to combine the equations \eq{2.1} -- \eq{2.3} with the
evolution equation 
\begin{equation}\label{E3.5}
S_t(t,x) = - f\big( {\cal F}(t,x)\big) |\na_x S(t,x)|,   
\end{equation}
with the driving force ${\cal F}$ defined by \eq{3.1} and with a given
linear or nonlinear function $f:\R \ra \R$. If we compare \eq{3.4} and
\eq{3.5}, then we see that the propagation speed of the diffuse
interface defined by \eq{3.5} is equal to $s=f\big({\cal F}(t,x)\big)$,
whence the kinetic relation is given by $f$ and can be read off
directly from the evolution equation \eq{3.5}. From \eq{2.8} we
immediately see that every solution $(u,T,S)$ of the equations
\eq{2.1}, \eq{2.2}, \eq{3.5} satisfies the Clausius-Duhem inequality
\eq{2.6} if $f$ satisfies $r \cdot f(r) \geq 0$ for all $r \in \R$.
The evolution equation \eq{3.5} has therefore the desired properties.

\eq{3.5} has the form of a Hamilton-Jacobi equation. However, if one
inserts the definition \eq{3.1} of ${\cal F}$ into \eq{3.5}, one
obtains the phase field equation
\begin{equation}\label{E3.6}
S_t = - f\big(\pa_S {\sf W} + \hat{\psi}'(S) - \Da_x S \big) |\na_x
  S|,  
\end{equation}
which is degenerate parabolic. \eq{3.6} has therefore mixed
hyperbolic--parabolic properties. This is why we call \eq{3.6} hybrid
phase field equation.

\section{The Allen-Cahn and the hybrid models}\label{S4}

We have now two different phase field models for the evolution of
phase interfaces in solids: If we combine the equations \eq{2.1},
\eq{2.2} with the phase field equation \eq{2.9} of Allen-Cahn type we
obtain the system  
\begin{eqnarray}
-\div_x\, T &=& {\sf b}, \label{E4.1}
\\
T &=& D\big(\ve(\na_x u ) - \ov{\ve}S \big),\label{E4.2}
\\
\pa_t S &=& - \frac{c}{(\mu\la)^{1/2}} \Big(\pa_S {\sf W} \big(\ve(\na_x
  u ), S \big) + \frac{1}{\mu^{1/2}} \hat{\psi}'(S) -
  \mu^{1/2} \la \Da_x S \Big), \label{E4.3} 
\end{eqnarray}
which must be solved in the domain $[0,\infty)\ti \Om$. As boundary
and initial conditions we choose, for example, 
\begin{alignat}{2}
u(t,x) &= {\sf U}(t,x), &\qquad& (t,x) \in [0,\infty)\ti \pa\Om,  
  \label{E4.4}
\\
\pa_{n_{\pa\Om}} S(t,x) &= 0, && (t,x) \in [0,\infty)
  \ti \pa\Om,  \label{E4.5}
\\ 
S(0,x) &= {\sf S}(x), && x \in \Om.  \label{E4.6}
\end{alignat}
To obtain \eq{4.3} from \eq{2.9} we specialized the function $f$
in \eq{2.9} to be $f(r) = cr$ with a positive constant $c$ and we
introduced two scaling parameters $\mu > 0$ and $\la > 0$, whose
meaning will become clear later. To have a short name, we call the
system \eq{4.1} -- \eq{4.3} the Allen-Cahn phase field model.

The second model is obtained by combination of \eq{2.1}, \eq{2.2} with
the hybrid phase field equation \eq{3.6}. If we specialize the
function $f$ in \eq{3.6} to be $f(r) = cr$ with a constant $c>0$ and
introduce a scaling parameter $\nu > 0$, the resulting system is 
\begin{eqnarray}
-\div_x\, T &=& {\sf b}, \label{E4.7}
\\
T &=& D\big(\ve(\na_x u ) - \ov{\ve}S \big),\label{E4.8}
\\
\pa_t S &=& -c \Big(\pa_S {\sf W} \big(\ve(\na_x
  u ), S \big) + \hat{\psi}'(S) - \nu \Da_x S \Big)|\na_x
  S|.   \label{E4.9} 
\end{eqnarray}
These equations must be solved in the domain $[0,\infty)\ti \Om$. For
the boundary and initial conditions we can again take \eq{4.4} --
\eq{4.6}. We call the system \eq{4.7} -- \eq{4.9} the hybrid phase
field model.

Several questions arise immediately. \eq{4.9} is a quasilinear,
degenerate parabolic equation.  Little is known about equations of the
form \eq{4.9}. The first question therefore concerns existence and
uniqueness of solutions to the system \eq{4.7} -- \eq{4.9}. Moreover,
if solutions $(u,T,S)$ exist, does the function $S$ have the
properties required from an order parameter? If both questions can be
answered positively, what is then the difference between the
Allen-Cahn model and the hybrid model? We have studied these questions
in recent years. To the first two questions only partial answers can
be given, whereas the answer to the third question is quite well
known.

In \cite{SIAM2006} it is proved that weak solutions of the hybrid
model \eq{4.7} -- \eq{4.9}, \eq{4.4} -- \eq{4.6} exist in the case of
one space dimension. The proof is based on the observation that the
one-dimensional version of the evolution equation \eq{4.9} has some
monotonicity properties. In higher space dimensions no rigorous
existence proof is available.  We must therefore rely on extensive
numerical tests and on formal asymptotic analysis. The numerical test
computations seem to indicate quite clearly, that solutions $(u,T,S)$
exist and that the function $S$ in these solutions has the properties
required from an order parameter.  In fact, the test computations
converge in higher space dimensions better then in one space
dimension. A part of the test computations is documented in
\cite{JElast2013}.

The last question on the difference of the models is answered in the
remainder of this paper. Of course, to answer the question we need to
have more information on the properties of the models. This
information is collected in Sections~\ref{S5} and \ref{S6}. The
information is obtained by asymptotic analysis of the models, more
precisely by construction of approximate solutions to the Allen-Cahn
and the hybrid models. The answer to the comparison question is
finally given in Section~\ref{S7}.

\section{Model error and asymptotics}\label{S5}

To compare the models we need to define what we understand under the
model error. In this section we first give this definition and
subsequently state in Theorems~\ref{T5.2} and \ref{T5.3} some results
on approximate solutions, which have been obtained in
\cite{JElast2013} and \cite{AC2015}.

To define the model error we must first specify the type of material
interfaces, which we want to model. Of great current interest are
phase interfaces in functional materials. Very often such interfaces
are thin and consist only of a few atomic layers. A large number of
phase field models to simulate the time evolution of such interfaces
have been devised and more are developed. It is therefore of interest
to study how well the Allen-Cahn and the hybrid models are adapted to
the simulation of thin interfaces in solids. More precisely, it is of
interest to study how large the difference between the propagation
speed of a thin phase interface in the real material and of the
interface in the respective phase field model is. This difference is
the model error.

To give a precise definition of the model error, we must approximately
know the propagation speed of the real phase interface. For very thin
interfaces mathematical models with sharp interface are appropriate.
We therefore base the following considerations on the hypothesis that
the propagation speed of the interface in the sharp interface model is
a good approximation to the propagation speed of the interface in the
real material. The model error of a phase field model is then the
difference of the propagation speed of the sharp interface and the
propagation speed of the diffuse interface in the phase field model.

To formulate the sharp interface model to be used we must introduce
some notations. 
The asymptotic solution is constructed in the bounded domain
\[
Q = [t_1,t_2] \ti \Om,
\]
where $0\leq t_1 < t_2 < \infty$ are given times. $\Gm(t)$ denotes the
sharp interface at time $t$. We assume that the phase sets $\gm(t)$,
$\gm'(t)$ introduced in Section~\ref{S2} are open, disjoint subsets of
$\Om$, whose common boundary is $\Gm(t)$, such that $\Om = \gm(t) \cup
\gm'(t) \cup \Gm(t)$.  We set
\begin{eqnarray*}
\Gm &=& \{ (t,x) \in Q \mid x \in \Gm(t),\ t_1 \leq t
\leq t_2 \}, \\
\gm &=& \{ (t,x) \in Q \mid x \in \gm(t),\ t_1 \leq t
  \leq t_2 \}, \\
\gm' &=& \{ (t,x) \in Q \mid x \in \gm'(t),\ t_1
  \leq t \leq t_2 \}.
\end{eqnarray*}
Let 
\[
n : \Gm \ra \R^3
\]
be the continuous vector field, for which $n(t,x)$ is the unit normal
vector to $\Gm(t)$ at $x \in \Gm(t)$, which points into
the domain $\gm'(t)$. For a function $w$ defined in a
neighborhood of $\Gm$ and $(t,x) \in \Gm$ we set
\begin{eqnarray*}
w^{(\pm)}(t,x) &=& \lim_{\xi \searrow 0} w\big(t,x \pm n(t,x)
  \xi\big),\\
{[w]}(t,x) &=& w^{(+)}(t,x) - w^{(-)}(t,x), \\
\langle w \rangle (t,x) &=& \frac12 \Big(w^{(+)}(t,x) + w^{(-)}(t,x)
  \Big).  
\end{eqnarray*}
Now we can formulate the sharp interface model. Let $\hS:Q \ra
\{0,1\}$ be a piecewise constant function, which only takes the values
$0$ and $1$ with a jump across $\Gm$, such that
\[
\gm(t) = \{ x \in \Om \mid \hS(t,x) = 0 \}, \quad \gm'(t) = \{ x \in
\Om \mid \hS(t,x) = 1 \}. 
\] 
The sharp interface model consists of a transmission problem for the
elasticity equations and of a kinetic relation. The transmission
problem is given by
\begin{eqnarray}
\label{E5.1}
-\div_x \hat{T} &=& {\sf b},
\\
\label{E5.2}
\hat{T} &=& D\big( \ve(\na_x \hat{u}) - \ov{\ve} \hat{S}\big),
\\
\label{E5.3}
{[\hat{u}]} &=& 0,
\\
\label{E5.4}
{[\hat{T}]}n &=& 0,
\\
\label{E5.5}
\hu(t)\rain{\pa\Om} &=& {\sf U}(t).	
\end{eqnarray}
To determine the kinetic relation we proceed as in Section~\ref{S2}.
We use the Clausius-Duhem inequality
\begin{equation}\label{E5.6}
\pa_t \psi_{\rm sharp} + \div_x\, q_{\rm sharp} \leq \hu_t \cdot {\sf
  b}, 
\end{equation}
with the free enery and the flux
\begin{eqnarray}
\psi_{\rm sharp}\big(\ve(\na_x \hat{u}), \hat{S}\big) &=&  {\sf W}
\big(\ve(\na_x \hat{u}), \hat{S}\big) + \lambda^{1/2} c_1 
 \int_{\Gamma(t)}\,{\rm d} \sigma, \label{E5.7}
\\
q_{\rm sharp} ( \hT, \hS ) &=& - \hT \cdot \hu_t \,, \nn 
\end{eqnarray}
where $c_1 \geq 0$ is an arbitrarily chosen constant. The last term on
the right hand side of \eq{5.7} is the interface energy, hence
$\la^{1/2} c_1$ is the interface energy density. It is well known that
if $(\hu,\hT)$ is a solution of the transmission problem \eq{5.1} --
\eq{5.5} and if the interface $\Gm(t)$ in this problem moves with the
normal speed $s_{\rm sharp}(t,x)$ at $x \in \Gm(t)$, then the
Clausius-Duhem inequality \eq{5.6} holds if and only if the inequality
\begin{equation}\label{E5.8}
s_{\rm sharp}(t,x)\cdot \Big( {-} \ov{\ve}: \langle \hT \rangle(t,x) +
  \la^{1/2} c_1  \ka_\Gm(t,x) \Big) \geq 0 
\end{equation}
is satisfied at every point $x \in \Gm(t)$. Here $\ka_\Gm(t,x)$
denotes twice the mean curvature of the surface $\Gm(t)$ at $x \in
\Gm(t)$.

A proof of this well known result is given in \cite{Al00}, however
only for the case where $c_1 = 0$ in \eq{5.7}. The proof can be
readily generalized to the case $c_1 > 0$.

A simple linear kinetic relation, for which \eq{5.8} obviously
holds, is 
\begin{equation}\label{E5.9}
s_{\rm sharp} = \hat{c}\, \big( - \ov{\ve}: \langle \hT \rangle +
  \la^{1/2} c_1  \ka_\Gm \big), 
\end{equation}
with a positive constant $\hat{c}$. The sharp interface problem thus
consists of the transmission problem \eq{5.1} -- \eq{5.5} combined
with the kinetic relation \eq{5.9}.

We can now define the model error. To this end note that solutions of
the Allen-Cahn model depend on the parameters $\mu$ and $\la$, whereas
solutions of the hybrid model depend on the parameter $\nu$. Therefore
we record these parameters in the notation. For a solution 
$(u^{(\mu\la)}_{\rm AC},T^{(\mu\la)}_{\rm AC},S^{(\mu\la)}_{\rm AC})$
of the Allen-Cahn model and for a solution $(u^{(\nu)}_{\rm
  hyb},T^{(\nu)}_{\rm hyb}, S^{(\nu)}_{\rm hyb})$ of the hybrid model
consider the level sets
\[
\Gamma_{\rm AC}^{(\mu\la)}  = \Big\{ (t,x) \in Q \Bigm|
  S_{\rm AC}^{(\mu\la)}(t,x)=\frac{1}{2}\Big\}, 
\quad
\Gamma_{\rm hyb}^{(\nu)}  = \Big\{ (t,x) \in Q \Bigm|
  S_{\rm hyb}^{(\nu)}(t,x)=\frac{1}{2}\Big\}. 
\] 
Let $s_{\rm AC}^{(\mu\la)}(t,x)$ denote the normal speed of
$\Gamma_{\rm AC}^{(\mu\la)}(t)$ at $x \in \Gamma_{\rm
  AC}^{(\mu\la)}(t)$, and let $s_{\rm hyb}^{(\nu)}(t,x)$ denote the
normal speed of $\Gamma_{\rm hyb}^{(\nu)}(t)$ at $x \in \Gamma_{\rm
  hyb}^{(\nu)}(t)$. These normal speeds are approximately equal to the
propagation speeds of the diffuse phase interfaces defined by the
solutions of the Allen-Cahn and hybrid models. 

Let $t \in [t_1,t_2]$ be a given, fixed number. As initial conditions
for the sharp interface problem we can choose
\[
\Gm(t) = \Gamma_{\rm AC}^{(\mu\la)}(t),\quad
\mbox{or}\quad \Gm(t) = \Gamma_{\rm hyb}^{(\nu)}(t). 
\]

\begin{tdefi}\label{D5.1}
We call the functions ${\cal E}^{(\mu\la)}(t):\Gm(t) \ra
\R$ and ${\cal E}^{(\nu)}(t):\Gm(t) \ra
\R$, respectively, which are defined by  
\begin{eqnarray}
{\cal E}^{(\mu\la)}(t) &=& s_{\rm AC}^{(\mu\la)}(t) - s_{\rm
  sharp}(t), \qquad \mbox{if } \Gm(t) = \Gamma_{\rm
  AC}^{(\mu\la)}(t), \label{E5.10} 
\\
{\cal E}^{(\nu)}(t) &=& s_{\rm hyb}^{(\nu)}(t) - s_{\rm
  sharp}(t), \qquad \mbox{if } \Gm(t) = \Gamma_{\rm
  hyb}^{(\nu)}(t), \label{E5.11} 
\end{eqnarray}
the error of the Allen-Cahn model or the error of the hybrid model at
time $t$, respectively. 
\end{tdefi}
We next state some results for the Allen-Cahn and hybrid models
obtained by asymptotic analysis. 

By $B_{\rm AC}^{(\mu\la)}>0$ and $B_{\rm hyb}^{(\nu)}>0$ we denote the
widths of the diffuse interfaces defined by the order parameter in
solutions of the Allen-Cahn model and by the order parameter in
solutions of the hybrid model. Here we do not define the interface
width precisely. If $S$ is an order parameter, one could define the
interface width to be the maximal distance between the level surfaces
$\{ x \in \Om \mid S(t,x) = 0.1\}$ and $\{ x \in \Om \mid S(t,x) =
0.9\}$, for example. We are interested in the limits $\mu \ra 0$, $\la
\ra 0$, $\nu \ra 0$ and assume therefore that $\mu \in (0,\mu_0]$,
$\la \in (0,\la_0]$, $\nu \in (0,\nu_0]$, with suitably chosen fixed
constants $\mu_0,\la_0,\nu_0 > 0$.

\begin{theo}\label{T5.2}
Let $(u^{(\mu\la)}_{\rm AC},T^{(\mu\la)}_{\rm AC},S^{(\mu\la)}_{\rm
  AC})$ be a solution of the Allen-Cahn model \eq{4.1} -- \eq{4.5},
let $t \in [t_1,t_2]$ be a given time, and
let $(\hu(t),\hT(t))$ be the solution of the transmission problem
\eq{5.1} -- \eq{5.5} with the interface given by 
$\Gm(t) = \Gm_{\rm AC}^{(\mu\la)}(t)$. Then 
\begin{equation}\label{E5.12}
s_{\rm AC}^{(\mu\la)}(t,x) = s_0(t,x) + \mu^{1/2} \big( s_{10}(t,x) +
 \la^{1/2} s_{11}(t,x) \big) + \mu^{1/2} R_{\rm AC} (\mu,\la,t,x),
\end{equation}
where $s_0 = s_0\big(\Gm_{\rm AC}^{(\mu\la)}(t) \big)$, $s_{10} =
s_{10} \big(\Gm_{\rm AC}^{(\mu\la)}(t) \big)$ and $s_{11}\big(\Gm_{\rm
  AC}^{(\mu\la)}(t) \big)$ are nonlocal functions of $\Gm_{\rm
  AC}^{(\mu\la)}(t)$. In particular, we have  
\begin{equation}\label{E5.13}
s_0(t,x) = \frac{c}{c_1} \big( - \ov{\ve}: \langle 
    \hT \rangle (t,x) + \la^{1/2} c_1  \ka_\Gm(t,x) \big),
\end{equation}
with the constant
\begin{equation}\label{E5.14}
c_1 = \int_0^1 \sqrt{2\hat{\psi} (r)} dr.
\end{equation}
For the remainder term $R_{\rm AC} (\mu,\la,t,x)$ there is a function
$\mu \ra {\cal C}_{\cal E}(\mu)$ with the property that $\lim_{\mu \ra
  0}{\cal C}_{\cal E}(\mu) = 0$, such that for all $0 < \mu \leq
\mu_0$, $0 < \la \leq \la_0$ and all $(t,x) \in \Gm_{\rm
  AC}^{(\mu\la)}$ the inequality
\begin{equation}\label{E5.15}
|R_{\rm AC} (\mu,\la,t,x)| \leq {\cal C}_{\cal E}(\mu)
\end{equation} 
holds. Moreover, there is a constant $C_1 > 0$ such that for all $0 <
\mu \leq \mu_0$, $0 < \la \leq \la_0$  
\begin{equation}\label{E5.16}
B_{\rm AC}^{(\mu\la)} \leq C_1 (\mu\la)^{1/2}.
\end{equation}

\end{theo}
These results are contained in \cite{AC2015}. We stress here the fact,
that the results are obtained by formal asymptotic analysis. No
rigorous mathematical proof of these statements is given in
\cite{AC2015}. The asymptotic analysis with respect to $\mu \ra 0$
uses mathematical methods, which are standard in the analysis of phase
field models. This is different for the estimate \eq{5.15}, which says
that the remainder term $R_{\rm AC}$ tends to zero for $\mu \ra 0$,
uniformly with respect to $\la$. This uniformity 
estimate is obtained
by a second asymptotic analysis with respect to $\la \ra 0$. The
formal derivation of this estimate is a novelty introduced in
\cite{AC2015}.

\begin{theo}\label{T5.3}
Let $(u^{(\nu)}_{\rm hyb},T^{(\nu)}_{\rm hyb},S^{(\nu)}_{\rm
  hyb})$ be a solution of the hybrid model \eq{4.7} -- \eq{4.9},
\eq{4.4}, \eq{4.5}, let $t \in [t_1,t_2]$ be a given time, and
let $(\hu(t),\hT(t))$ be the solution of the transmission 
problem \eq{5.1} -- \eq{5.5} with the interface given by
$\Gm(t) = \Gm_{\rm hyb}^{(\nu)}(t)$. Then 
\begin{equation}\label{E5.17}
s_{\rm hyb}^{(\nu)}(t,x) =  c \Big( -\ov{\ve}: \langle \hT \rangle (t,x) +
  \nu^{1/2} R_{\rm hyb}(\nu,t,x) \Big),
\end{equation}
where $c>0$ is the constant from \eq{4.9}. For the remainder term
$R_{\rm hyb} (\nu,t,x)$ there is a constant $C_2$ such that for all
$0 < \nu \leq \nu_0$ and all $(t,x) \in \Gm_{\rm hyb}^{(\nu)}$ the
inequality 
\begin{equation}\label{E5.18}
| R_{\rm hyb} (\nu,t,x) | \leq C_2
\end{equation}
holds. Moreover, there is a constant $C_3 > 0$ such that for all $0 <
\nu \leq \nu_0$   
\begin{equation}\label{E5.19}
B_{\rm hyb}^{(\nu)} \leq C_3 \nu^{1/2}.
\end{equation}

\end{theo}
These results are obtained in \cite{JElast2013}, again by formal
asymptotic analysis.

\section{Characteristic equations}\label{S6}

From the results on the asymptotic behavior of the models stated in
Theorems~\ref{T5.2} and \ref{T5.3} we derive in this section for both
models some relations between parameters of the models. We call
these relations the characteristic relations of the models. The
comparison of the models in Section~\ref{S7} is based on these
relations.

We first consider the Allen-Cahn model. For $c_1$ in the free energy
\eq{5.7} we choose the value given by \eq{5.14}, With this value we
adapt the interface energy density $\la^{1/2} c_1$ to the value in the
real material by varying $\la$. In \eq{4.3} we choose $c =
\hat{c}c_1$. By \eq{5.9} and \eq{5.13} we then have 
\[ 
s_0 = s_{\rm sharp}\,, 
\] 
hence \eq{5.10} and \eq{5.12} together imply 
\begin{equation}\label{E6.1}
{\cal E}^{(\mu\la)} = s_{\rm AC}^{(\mu\la)} - s_0 = \mu^{1/2} ( s_{10}
  + \la^{1/2} s_{11} ) + \mu^{1/2} R_{\rm AC}\,.  
\end{equation}
This equation and \eq{5.15} together yield  
\begin{equation}\label{E5.20}
|{\cal E}^{(\mu\la)}| \leq C \mu^{1/2}, 
\end{equation}
with a constant $C$, which can be chosen independently of $\la$. By
this inequality, $\mu^{1/2}$ controls the model error. Therefore we
write $F = \mu^{1/2}$ and call $F$ the error parameter. Moreover,
since $\la^{1/2} c_1$ is the interface energy density, we call $E =
\la^{1/2}$ the interface energy parameter. Also, since by \eq{5.16}
the interface width is bounded by a constant, which is proportional to
$(\mu\la)^{1/2}$, we call ${\cal W} = (\mu\la)^{1/2}$ the interface
width parameter. These three parameters and the propagation speed
$s_{\rm AC} = s_{\rm AC}^{(\mu\la)}$ are connected by the fundamental
relations 
\begin{eqnarray}
{\cal W} &=& EF, \label{E5.21}
\\
s_{\rm AC} &=& \hat{c}\, ( - \ov{\ve}: \langle \hT \rangle  + c_1
  \ka_\Gm E ) + {\cal E}[E,F],  \label{E5.22} 
\\
|{\cal E}[E,F]| &\leq& C F,   \label{E5.23} 
\end{eqnarray}
where we use the notation ${\cal E}[E,F] = {\cal E}^{(\mu\la)}$. The
first equation is an immediate consequence of the definition of the
parameters, the second is obtained by insertion of \eq{5.9} into
\eq{5.10}, and the last inequality is just a restatement of \eq{5.20}.

Now assume that we want to use a phase field model to numerically
simulate the propagation of a phase interface. In such a simulation
the numerical effort is proportional to $h^{-p}$, where $h$ denotes
the grid spacing and where the power $p > 1$ depends on whether we
want to simulate a problem in $2$--d or in $3$--d and it depends on
the numerical scheme we use. In order for the simulation to be
precise, we must guarantee that the model error and the numerical
error are small. To make the numerical error small, we must choose the
grid spacing $h$ small enough to resolve the transition of the order
parameter across the interface, which means that we must choose $h <
{\cal W}$, hence we have $h^{-p} > {\cal W}^{-p}$. Therefore we see that the
numerical effort of a simulation based on a phase field model is
measured by the number ${\cal W}^{-p}$. We call the number
\[
e_{\rm num} = {\cal W}^{-p}  
\]
the parameter of numerical effort. For a simulation based on the
Allen-Cahn model we see from \eq{5.21} that the numerical
effort is  
\begin{equation}\label{E5.24}
e_{\rm num} = (EF)^{-p}.
\end{equation}
We call the relations \eq{5.21} -- \eq{5.24} characteristic relations
for the Allen-Cahn model.

Next we derive the characteristic relations for the hybrid model. In
the free energy \eq{5.7} we choose $c_1 = 0$, and in \eq{4.9} we set
$c = \hat{c}$. By \eq{5.9} and \eq{5.17} we then have
\begin{equation}\label{E5.25}
s^{(\nu)}_{\rm hyb} = s_{\rm sharp} + \nu^{1/2} \hat{c}\, R_{\rm hyb}.
\end{equation}
We insert this equation into \eq{5.11} and obtain for the model error 
\begin{equation}\label{E5.26}
{\cal E}^{(\nu)} = \hat{c} R_{\rm hyb} \, \nu^{1/2}.
\end{equation}
From this equation and from \eq{5.18} we infer that 
\begin{equation}\label{E5.27} 
| {\cal E}^{(\nu)}| = \hat{c} |R_{\rm hyb}| \nu^{1/2} \leq C \nu^{1/2}.
\end{equation}
By this equation, $\nu^{1/2}$ controls the model error. In the case of
the hybrid model we therefore choose $F=\nu^{1/2}$ as the error
parameter. By \eq{5.19}, the interface width is bounded by a constant,
which is proportional to $\nu^{1/2}$, whence the interface width
parameter is ${\cal W} = \nu^{1/2}$. For the hybrid model we therefore
have the characteristic relations  
\begin{eqnarray}
{\cal W} &=& F, \label{E5.28}
\\
s_{\rm hyb} &=& - \hat{c}\: \ov{\ve}: \langle \hT \rangle  + {\cal 
  E}[F],  \label{E5.29}  
\\
|{\cal E}[F]| &\leq& C F,   \label{E5.30} 
\\
e_{\rm num} &=& F^{-p}, \label{E5.31}
\end{eqnarray}
where we used the notations $s_{\rm hyb} = s_{\rm hyb}^{(\nu)}$ and
${\cal E}[F] = {\cal E}^{(\nu)}$. The first of these relations follows
from the definitions of $F$ and ${\cal W}$, the second one is obtained by
combination of \eq{5.25} and \eq{5.26}, noting \eq{5.9}, the third one
is just a restatement of \eq{5.27}, and the last one follows from the
definition $e_{\rm num} = {\cal W}^{-p}$ of the parameter of numerical
effort and from \eq{5.28}.

\section{Comparison of the models, numerical efficiency}\label{S7}

From \eq{5.22} we see that the Allen-Cahn model can describe the
evolution of a phase interface with propagation speed $\hat{c}\, ( -
\ov{\ve}: \langle \hT \rangle + c_1 \ka_\Gm E )$, which by \eq{5.9} is
the propagation speed of an interface with interface energy density
$c_1 \la^{1/2} = c_1 E$. The interface energy density is always
positive, since we cannot set $\la = 0$ in the Allen-Cahn equation
\eq{4.3}. Varying of the parameter $E$ to adjust the interface energy
density does not change the model error; this error can be adjusted to
a desired value by choosing the parameter $F = \mu^{1/2}$ suitably.
Varying of $F$ does not change the interface energy density. From
\eq{5.24} we see that if the interface energy density parameter $E$ is
fixed, then the effort of a numerical simulation grows with $F^{-p}$,
where the power $p > 1$ depends on the numerical method employed and
on the space dimension of the problem, which we want to simulate.

From \eq{5.29} we see that the hybrid model, on the other hand, can
describe the evolution of a phase interface with propagation speed
$- \hat{c}\:  \ov{\ve}: \langle \hT \rangle$, which by \eq{5.9} is
the propagation speed of an interface with interface energy density
$c_1 \la^{1/2} = 0$. The model error can be adjusted to a desired
value by choosing the parameter $F = \nu^{1/2}$ suitably.  By
\eq{5.31}, also for this model the effort of a numerical simulation
grows with $F^{-p}$, where the power $p > 1$ depends on the numerical
method employed and on the space dimension of the problem, which we
want to simulate.

These observations suggest the following rule:
\\[1ex]
{\em Simulations of phase interfaces with positive interface energy
  density should be based on the Allen-Cahn model, simulations of
  interfaces with zero or small interface energy density should be
  based on the hybrid model.}
\\[1ex]
One can object to this rule by arguing that the Allen-Cahn model can
also be used to simulate interfaces with zero interface energy density
by choosing the interface energy density parameter positive, but very
small. However, because of the presence of the factor $E^{-p}$ in the
formula \eq{5.24} the numerical effort will become very large.

To be more specific, we consider an interface with vanishing interface
energy density, hence $c_1 \la^{1/2}=0$, which by \eq{5.9} means that
the propagation speed of the sharp interface is
\[
s_{\rm sharp} = - \hat{c}\: \ov{\ve} : \langle \hT \rangle.
\]
For the Allen-Cahn model it follows from this equation and from
\eq{5.22} that in this case the total model error, which we denote by
${\cal E}_{\rm total}$, is 
\[
{\cal E}_{\rm total} = s_{\rm AC} - s_{\rm sharp} = \hat{c}c_1\ka_\Gm
  E + {\cal E}[E,F] .  
\] 
This means that the term $\hat{c}c_1\ka_\Gm E$ is now part of the
total model error. 

If we prescribe the maximal value ${\cal E}_{\max}$ of the total model
error $|{\cal E}_{\rm total}|$, we must therefore choose the
parameters $E$ and $F$ such that
\begin{eqnarray}
 \hat{c}c_1 (\max_\Gm |\ka_\Gm|)E + \max_\Gm |{\cal E}[E,F]| &\leq&
 {\cal E}_{\max},  \label{E7.1}  
\\
EF &\overset{!}{=}& \max, \label{E7.2}
\end{eqnarray}
where the second condition is imposed by the requirement to make the
numerical effort $e_{\rm num} = (EF)^{-p}$ as small as possible. To
discuss this optimization problem, we assume first that the term
$s_{10}$ in the asymptotic expansion \eq{5.12} of the propagation
speed $s^{(\mu\la)}_{\rm AC}$ is not identically equal to zero. In
this case we conclude from \eq{6.1} and \eq{5.15} that
for sufficiently small $\la^{1/2} = E$ and for sufficiently small
$\mu^{1/2} = F$ the error ${\cal E}[E,F] = {\cal E}^{(\mu\la)}$
satisfies  
\[
\max_\Gm |{\cal E}[E,F]| \geq \frac12 (\max_\Gm |s_{10}|) \mu^{1/2} = 
  \frac12 (\max_\Gm |s_{10}|) F.
\] 
This inequality and \eq{7.1} imply that the solution $(E,F)$ of
the optimization problem \eq{7.1}, \eq{7.2} satisfies
\[
F \leq \frac{2}{\max\limits_\Gm |s_{10}|}  \max_\Gm |{\cal E}[E,F]| \leq
   \frac{2}{\max\limits_\Gm |s_{10}|}\, {\cal E}_{\max} 
\qquad \mbox{and} \qquad E \leq \frac{1}{\hat{c}c_1 \max\limits_\Gm
  |\ka_\Gm|}\, {\cal E}_{\max} .
\]
From this result we obtain
\begin{coro}\label{C7.1}
Let ${\cal E}_{\rm max}$ denote the total model error of the
Allen-Cahn model in the simulation of an interface without interface
energy.  If the term $s_{10}$ in the asymptotic expansion \eq{5.12}
of the propagation speed $s^{(\mu\la)}_{\rm AC}$ is not identically
equal to zero, then the interface width $B_{\rm AC}$ satisfies
\begin{equation}\label{E7.3}
B_{\rm AC} \leq C_1 EF \leq \frac{2C_1}{\hat{c} c_1 (\max\limits_\Gm
  |s_{10}|) ( \max\limits_\Gm |\ka_\Gm|) }\, {\cal E}_{\max}^2\,.  
\end{equation}
In a numerical simulation of an interface without interface energy
based on the Allen-Cahn model the parameter of numerical effort
satisfies
\begin{equation}\label{E7.4}
e_{\rm num} \geq \left(
  \frac{\hat{c} c_1  (\max\limits_\Gm |s_{10}|)  
  (\max\limits_\Gm |\ka_\Gm|) }{ 2\,{\cal E}_{\max}^2} \right)^p ,
\end{equation} 
with a power $p > 1$ depending on the space dimension and the
numerical method used. 
\end{coro}
For the hybrid model we have by \eq{5.29} and \eq{5.30} that ${\cal
  E}_{\max} = \max_\Gm | {\cal E}[F]| \leq C F$. From \eq{5.31} and
from \eq{7.4} we thus see that in a simulation of an interface without
interface energy or with small interface energy the numerical efforts
behave like
\begin{equation}\label{E7.5}
e_{\rm num}^{\rm hyb} \leq C {\cal E}_{\max}^{-p}\,,\qquad  e_{\rm
  num}^{\rm AC} \geq C {\cal E}_{\max}^{-2p}\,. 
\end{equation}
Since the time step in a simulation must be decreased when the grid
spacing $h$ in $x$--direction is decreased, the number $p$ can be
larger than $4$ in a three dimensional simulation. From \eq{7.5} we
thus see that the numerical effort for the Allen-Cahn model grows much
faster for the Allen-Cahn model than for the hybrid model when the
required accuracy is increased. This confirms the rule stated above
for the usage of both models in simulations.

This picture does not change essentially when the term $s_{10}$
vanishes identically. In this case the same considerations show that
instead of \eq{7.3} and \eq{7.4} we would have $B_{\rm AC} = O({\cal
  E}_{\rm max}^{3/2})$ and $e_{\rm num}^{\rm AC} \geq C {\cal E}_{\rm
  max}^{-\frac32 p}$, hence the numerical effort for the Allen-Cahn
model would still grow faster than for the hybrid model.  However, a
close investigation of the terms, which constitute $s_{10}$ and which
are computed in \cite{AC2015}, shows that only in very exceptional
situations one can expect that $s_{10}$ vanishes identically.




\end{document}